\documentclass{article}

\usepackage{neurips_2023}

\usepackage{cite}
\usepackage{amsmath,amssymb,amsfonts}
\usepackage{algorithmic}
\usepackage{graphics}
\usepackage{textcomp}
\usepackage{xcolor}
\usepackage{enumitem}

\usepackage{graphicx}
\usepackage{dblfloatfix}

\usepackage{amsmath}

\usepackage{graphicx}
\usepackage{subcaption}

\usepackage[utf8]{inputenc} % allow utf-8 input
\usepackage[T1]{fontenc}    % use 8-bit T1 fonts
\usepackage{hyperref}       % hyperlinks
\usepackage{url}            % simple URL typesetting
\usepackage{booktabs}       % professional-quality tables
\usepackage{amsfonts}       % blackboard math symbols
\usepackage{nicefrac}       % compact symbols for 1/2, etc.
\usepackage{microtype}      % microtypography
\usepackage{xcolor}         % colors

\title{Exploring MLOps Dynamics: An Experimental Analysis in a Real-World Machine Learning Project}

\author{Awadelrahman M. A. Ahmed \thanks{Awadelrahman M. A. A. is also a PhD fellow at the Department of Informatics, University of Oslo, Norway} \\
  Data Scientist at Amesto NextBridge AS\\
  Oslo, Norway\\
  \texttt{mlopswiki@mlopswiki.com} \\
}

\begin{document}

\maketitle

\begin{abstract}

	This article presents an experiment focused on optimizing the MLOps (Machine Learning Operations) process, a crucial aspect of efficiently implementing machine learning projects. The objective is to identify patterns and insights to enhance the MLOps workflow, considering its iterative and interdependent nature in real-world model development scenarios.
	
	The experiment involves a comprehensive MLOps workflow, covering essential phases like problem definition, data acquisition, data preparation, model development, model deployment, monitoring, management, scalability, and governance and compliance. Practical tips and recommendations are derived from the results, emphasizing proactive planning and continuous improvement for the MLOps workflow.
	
   The experimental investigation was strategically integrated within a real-world ML project which followed essential phases of the MLOps process in a production environment, handling large-scale structured data. A systematic tracking approach was employed to document revisits to specific phases from a main phase under focus, capturing the reasons for such revisits. By constructing a matrix to quantify the degree of overlap between phases, the study unveils the dynamic and iterative nature of the MLOps workflow.
   
	The resulting data provides visual representations of the MLOps process's interdependencies and iterative characteristics within the experimental framework, offering valuable insights for optimizing the workflow and making informed decisions in real-world scenarios. This analysis contributes to enhancing the efficiency and effectiveness of machine learning projects through an improved MLOps process.
	
\end{abstract}
\newpage
\tableofcontents
\newpage
\section{Introduction}

The field of Machine Learning Operations, commonly known as MLOps, plays a pivotal role in efficiently deploying machine learning models to deliver real-world value in businesses and industries. With its iterative and interdependent nature, MLOps presents both challenges and opportunities for organizations seeking to harness the full potential of AI-driven solutions. This article presents an experimental investigation aimed at gaining valuable insights into the dynamics of the MLOps process and its iterative nature, facilitating optimization strategies.

Inspired by the book "Designing Machine Learning Systems: An Iterative Process for Production-Ready Applications" by Chip Huyen [1], which inherently reflects the iterative aspect of MLOps, this work delves deeper into understanding the iterative patterns present in real-world model development scenarios. The primary objective of the experiment is to examine key requirement patterns that necessitate revisiting other phases when focusing on specific phases within the MLOps process. The experiment tracked essential phases of the MLOps process in a real-world ML project to quantify their overlap, providing valuable insights for optimization. 

It is important to note the crucial distinction between \textit{iterativeness} and \textit{repetitiveness}. Iterativeness involves repeating tasks for continuous improvement and refinement, while repetitiveness implies mindlessly repeating with no (or trivial) progress.  In the context of MLOps, the aim should be following an iterative approach, driving improvement at every repetition in the process. 

This article maybe of value for MLOps engineers as well as domain experts, as it provides a common understanding and helps set a mindset and expectations regarding the iterative nature of the MLOps process in the real world. It is important to note that while the specific patterns may vary across different projects and ML use cases, the common attribute is the iterative nature, which is likely to be present in similar contexts.

In the subsequent sections, we describe the methodology and delve into each MLOps phase, analyzing the interdependencies and iterative characteristics encountered during the experiment. Ultimately, this article concludes with a set of tips and recommendations derived from the experiment's findings, offering practical guidance to optimize each MLOps phase. By leveraging these insights, businesses can  navigate the iterative nature of MLOps toward harnessing the full potential of machine learning in their operations.

\section{Methodology}

The experiment was designed as a tracking process, integrated into a real-world ML project. The project encompassed essential phases of the MLOps process, including problem definition, data acquisition, data preparation, model development, model deployment, monitoring and management, scalability, governance and compliance, and communication. The primary objective of the tracking experiment is to quantify the degree of overlap between these phases within the primary workflow of the main project.

The project's execution took place in a production environment and spanned months, utilizing structured data of significant magnitude, comprising tens of millions of rows on a national scale. To ensure data privacy and confidentiality, specific details concerning the dataset cannot be disclosed.

Throughout the experiment, detailed records were maintained to track the principal reasons triggering revisits to other MLOps phases from the primary phase under focus. This allowed us to address critical inquiries, such as identifying the most frequently revisited phases and understanding the specific needs prompting these revisits. 

To delve deeper into the iterative and interdependent nature of the MLOps workflow, our experiment incorporated a systematic tracking approach. We constructed a matrix to quantify the degree of overlap between the primary workflow phases. Detailed records documented each occurrence of revisiting a particular phase from a main phase under focus, along with the reasons for such revisits. For instance, if during model development, we revisited the data preparation phase 20\% of the time, the matrix displayed a value of 0.2 at the intersection between "model  development" and "data preparation".

The resulting data offered visual representations of the inter-dependencies and iterative nature of the MLOps process within the experimental framework, providing valuable insights into the dynamics of the workflow. These findings contribute to a deeper understanding of the iterative characteristics and guide the optimization of the MLOps process in real-world machine learning projects.

\section{MLOps Workflow}
Nine phases were defined to be the main focus each time, based on which reservists to other phases were tracked. These main phases are:

\subsection{Problem Definition}
This phase involves concretely defining the problem and mapping it to the business objectives. It requires collaboration with subject matter experts (SMEs) to ensure a clear understanding of the problem statement and its relevance in the industrial context. 

The main focus of this phase is to answer the question: \textit{"What is the problem to be addressed, and how does it align with the business objectives?"}.

\subsection{Data Acquisition}
The data acquisition phase involves obtaining access to the required data for model development. This includes identifying data sources, acquiring data from internal or external repositories, and ensuring data availability and permissions.  

The main focus of this phase is to answer the question: \textit{"How can we access the necessary data for model development?"}.

\subsection{Data Preparation}
Data preparation encompasses activities such as data cleaning, data integration, data transformation, and feature engineering. This phase aimes to prepare the data in a suitable format for analysis and model development. 

So, the question of this phase is : \textit{"How to transform data to a suitable format for analysis and model development?"}.

\subsection{Model Development}
The model development phase involves training and building the predictive model using the prepared data. It includes selecting appropriate algorithms, tuning model parameters, and evaluating model performance to optimize the model's predictive capabilities. 

The question of this phase is : \textit{"How to build an effective predictive model?"}.

\subsection{Model Deployment}
This is the process of deploying the trained model into a production environment, making it available for use by stakeholders or end-users. This phase involves considerations such as model packaging, integration with existing systems, and setting up mechanisms for model serving and inference. 

The main question for this phase is:\textit{ "How to make the trained model available for stakeholders and end-users?"}.

\subsection{Monitoring and Management}
Here is where we start focusing on monitoring the deployed model's performance, tracking key metrics, and addressing any issues or anomalies that arise. It includes implementing monitoring systems, conducting periodic model evaluations, and ensuring ongoing model accuracy and reliability. 

The main question to answer in this phase: \textit{"How to ensure continuous model performance and reliability while addressing potential issues?"}.

\subsection{Scalability}

This phase focuses on assessing and enhancing the model's scalability to handle larger datasets or increased usage. It involves optimizing computational resources, evaluating infrastructure requirements, and implementing strategies to accommodate growth and increasing demands. 

The main question for this phase is: \textit{"How to optimize computational resources and infrastructure to accommodate growth and increasing demands?"}.

\subsection{Governance and Compliance}
Governance and compliance refers to adhering to ethical considerations, data privacy regulations, and industry-specific compliance requirements throughout the MLOps process. It includes implementing data governance practices, ensuring data security, and maintaining compliance with relevant policies and regulations. 

The main question to answer in this phase: \textit{"How to ensure ethical and compliant practices throughout the whole process?"}.

\subsection{Communication}
The communication phase, though not traditionally considered a separate phase, was deliberately highlighted in our workflow due to its critical role. The communication phase involves effectively communicating findings, insights, and outcomes of the model development process to stakeholders and decision-makers. 

The main question of this phase: \textit{"How to engage stakeholders and enable informed decision-making based on the model's results?"}.

\paragraph{Note}

Theoretically, these phases  can be perceived as sequential, focusing on their main questions, one question at a time. However, practical implementation involves iterative processes, re-visiting (backtracking) previous phases or proactively engaging (forward-thinking) with future phases. In that perspective, our experiment aims to identify reasons driving such iterations, streamlining the MLOps process for greater efficiency in real-world settings as we discuss in the next section.

\newpage

\section{Results \& Discussion}

In this section, we explore the main scenarios prompting transitions between MLOps phases. Each subsection focuses on a key phase, discussing revisited past phases and proactively approached future phases. Our analysis aims to optimize the MLOps process by addressing evolving needs. For each transition we provide optimization tips and actionable recommendations for improved, more efficient MLOps process.

\subsection{Results Overview}

\begin{figure}[htpb]
	\centering
	\includegraphics[width= \textwidth]{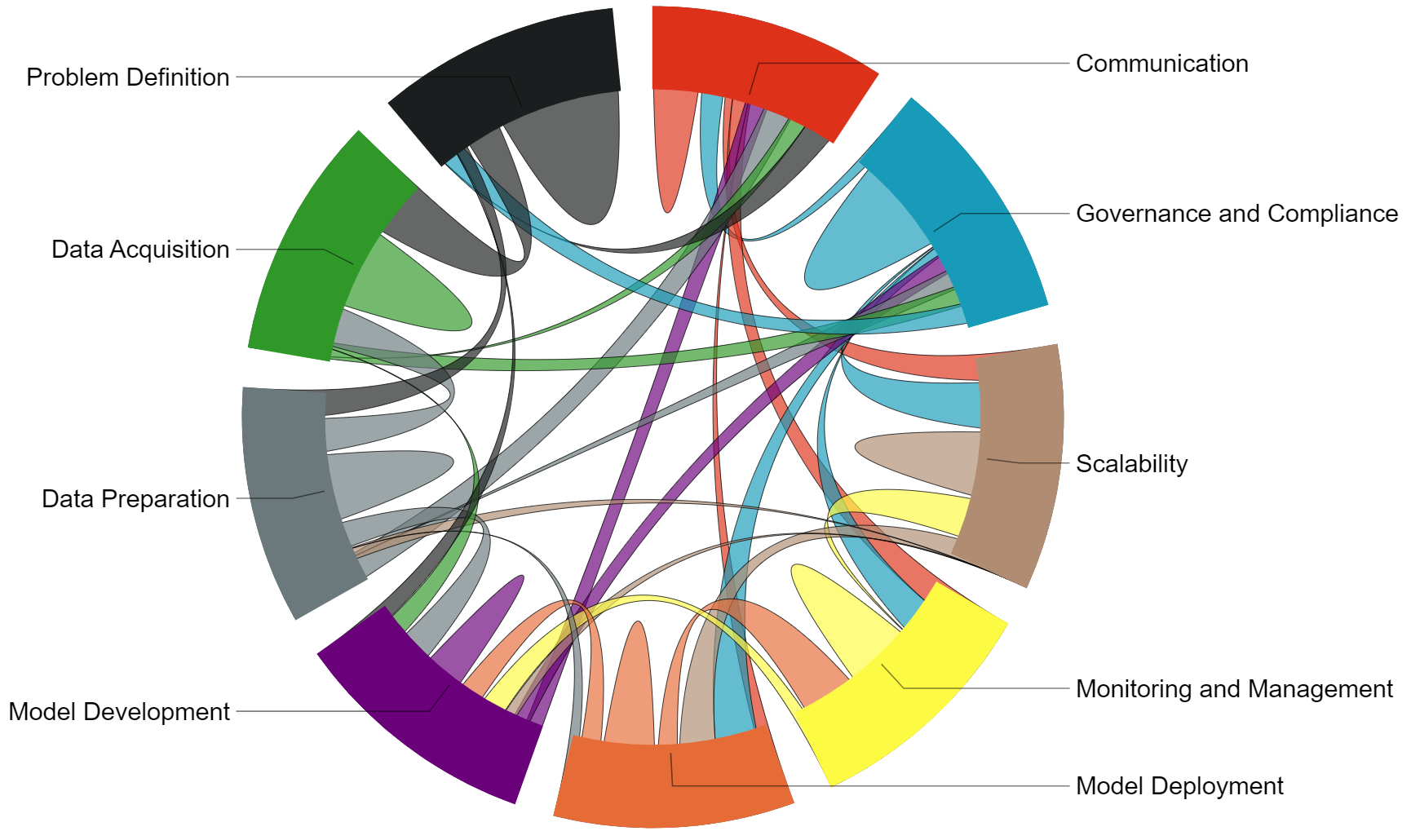}
	\centering
	\caption{Visualization of MLOps inter-dependencies in a real-world project}\label{sank}
\end{figure}

To get an overview of the dynamic nature of the process, we can visualize the Sankey Chord Diagram in Figure \ref{sank} which shows an overall picture of the iterative process between various phases in the experimented MLOps lifecycle. This shows how each phase may interact and influence other phases throughout the lifecycle. The flows in the diagram depict the movement of considerations, tasks, and feedback between phases, highlighting the need for revisiting certain phases based on the requirements and insights obtained in subsequent phases or proactively performing some tasks related to future phases.

\newpage
\subsection{Main Focus: Problem Definition}
\begin{figure}[htpb]
	
	\begin{subfigure}{\textwidth}
			\centering
	\includegraphics[width=0.9\textwidth]{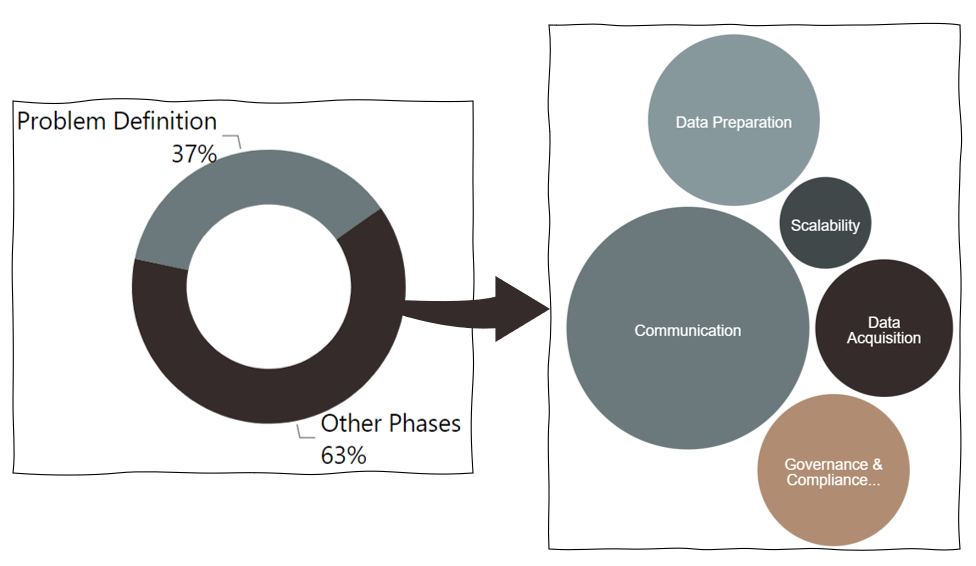}
	\caption{}\label{pd1}
	\end{subfigure}

	\begin{subfigure}{\textwidth}
			\centering
	\includegraphics[width=0.8\textwidth]{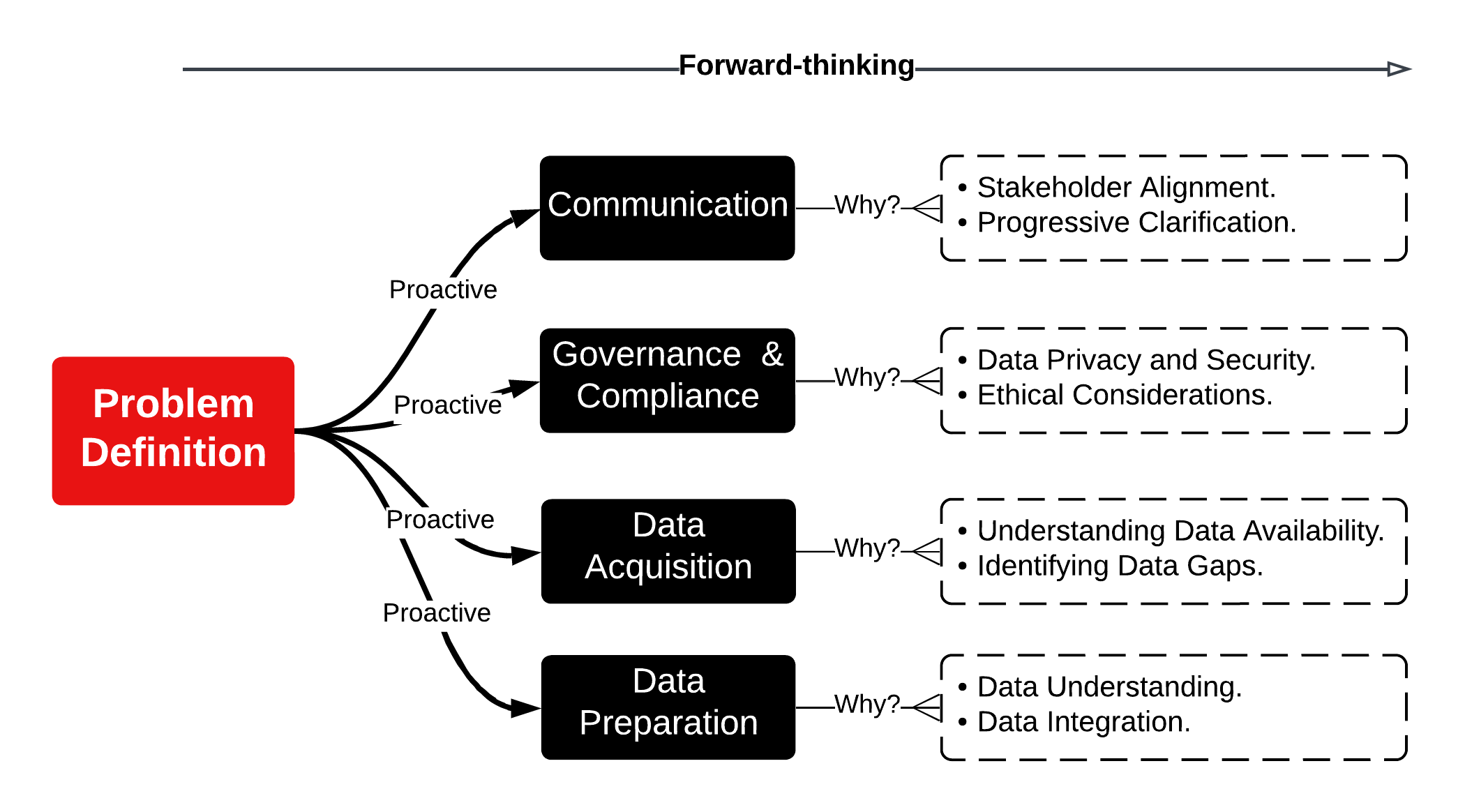}
	\caption{}\label{pd2}
	\end{subfigure}
		\centering
	\caption{Main Focus: Problem Definition  (a) The factual focus distribution when Problem Definition phase is planned to be the main focus  (b) Problem Definition phase inter-dependencies and main scenarios that trigger focus shift } \label{pd}

\end{figure}

When focusing the problem definition phase, we observed a proactive inclination towards four future phases with varying degrees of significance. The phases of communication, governance and compliance, data acquisition, and data preparation stood out with higher weights in terms of consideration. While other phases, such as scalability, were also considered, they held relatively lower significance, see Figure \ref{pd}.

\subsubsection{Forward-thinking: Communication}
In the context of the problem definition phase, communication emerges as a central activity that is naturally interlaced with the process, rather than being perceived as a separate phase that requires independent visitation. In our particular case, the significance of communication during the problem definition phase was emphasized by two main scenarios:

\begin{enumerate}
	
	\item \textit{Stakeholder Alignment}: Communication during the problem definition phase played a pivotal role in achieving stakeholder alignment. Engaging with stakeholders, we discussed essential questions regarding business challenges, domain-specific considerations, desired outcomes, and compliance.
	
	\item  \textit{Progressive Clarification}: During our experiment, we observed that the problem definition phase was characterized by progressive clarification rather than being accomplished in a single communication occasion. We will see later that even when we were sure the problem is defined, we come back to this phase from very advanced phases. This notion of progressive clarification helped in refining and evolving our understanding of the problem.
\end{enumerate}

\subsubsection{Forward-thinking: Data Acquisition}
Two primary reasons prompt proactive consideration of data acquisition during the problem definition phase:

\begin{enumerate}
	
	\item \textit{Understanding Data Availability}: During the problem definition phase, discussions about data acquisition often arise to understand the availability of data relevant to the defined problem. This aims to determine whether the required data is accessible and sufficient for addressing the problem at hand. This could involve checking existing databases, collaborating with data engineers, or exploring external data sources to determine if the required data can be obtained.
	
	\item  \textit{Identifying Data Gaps}: During the problem definition phase, it is possible to identify critical data requirements that are essential for gaining insights and addressing the defined problem. However, if this required data is missing or incomplete within the existing dataset, discussing data acquisition phase becomes necessary.
\end{enumerate}

\subsubsection{Forward-thinking: Data Preparation}

Two main examples of why data preparation phase is visited while defining the problem:
\begin{enumerate}
	\item  \textit{Data Understanding}:  Exploring the data preparation phase during problem definition is beneficial if not crucial. By understanding the data's characteristics, quality, and limitations, we can assess its suitability for the problem at hand. This allows us to identify potential issues, biases, or gaps that may impact model performance. 
	
	\item   \textit{Data Integration}:  In some cases, the problem definition phase may reveal that additional data sources need to be incorporated to enhance the predictive power of the model. This prompts a need to integrate external datasets into the existing data. For instance, you may identify the need to integrate customer satisfaction survey data or social media sentiment analysis data to enrich the information available for prediction. This integration task would be addressed within the data preparation phase.
\end{enumerate}

\subsubsection{Forward-thinking: Governance and Compliance}
During the problem definition phase, two key occasions emerged where proactive consideration of the governance and compliance phase took place:

\begin{enumerate}
	
	\item \textit{Data Privacy and Security}:  Within the context of defining the problem, data privacy naturally arises as a significant consideration. We engage in discussions and assessments to ensure compliance with relevant regulations, such as GDPR or HIPAA. That may involve implementing anonymization techniques and adhering to industry-specific regulations.
	
	\item  \textit{Ethical Considerations}: Another significant aspect during the problem definition phase is the exploration of ethical implications. We are expected to proactively discuss potential biases, fairness, and responsible AI practices to address ethical concerns.
\end{enumerate}

\subsubsection {Recommendations for Optimizing the Problem Definition Phase} 

Given those scenarios, to optimize the problem definition phase, we recommend:

\begin{enumerate}
	
	\item Embracing a mindset of continuous engagement and communication with stakeholders. Recognizing that the problem may not be fully clear at the beginning and that it can (will) evolve over time. 
	\item Actively seeking input from stakeholders ensures alignment and relevance, while documenting the evolving understanding of the problem facilitates clarity and informed decision-making. 
	\item Regularly validating and verifying the problem statement maintains its accuracy and effectiveness.

	\item  Integrating ethical considerations into the problem definition phase, discussing potential biases and ensuring responsible AI practices. For instance, the problem statement can be something like: \textit{"Developing an ethical recommender system that takes into account user privacy, fairness, and responsible artificial intelligence practices"}.
	\item Engaging with data privacy and security experts for guidance and best practices.
	
	\item  Conducting a comprehensive assessment of data availability early on in the problem definition phase. Collaborating with relevant stakeholders, data engineers, and domain experts to identify and access the necessary data sources.

	\item   If data from multiple sources or formats need to be integrated, identify and address potential challenges early on. Develop strategies for merging, aligning, or harmonizing diverse datasets to create a unified and coherent dataset for analysis.
	
\end{enumerate}

\newpage
\subsection{ Main Focus: Data Acquisition}
\begin{figure}[htp]
	
	\begin{subfigure}{\textwidth}
			\centering
		\includegraphics[width=0.8\textwidth]{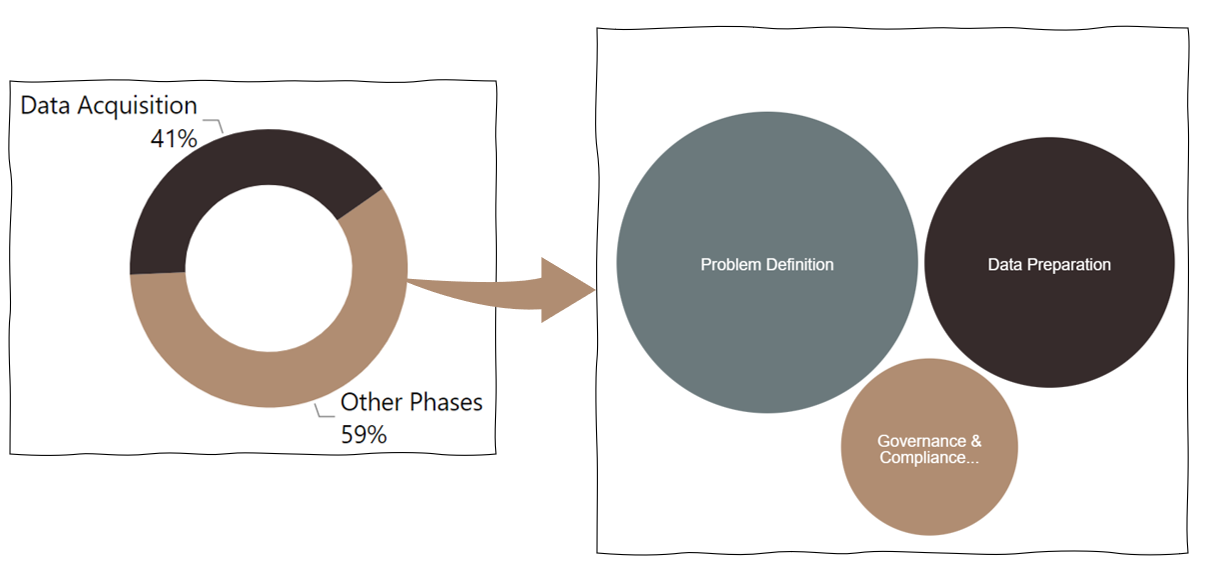}
		\caption{}\label{da1}
	\end{subfigure}
	
	\bigskip
	
	\begin{subfigure}{\textwidth}
			\centering
		\includegraphics[width= \textwidth]{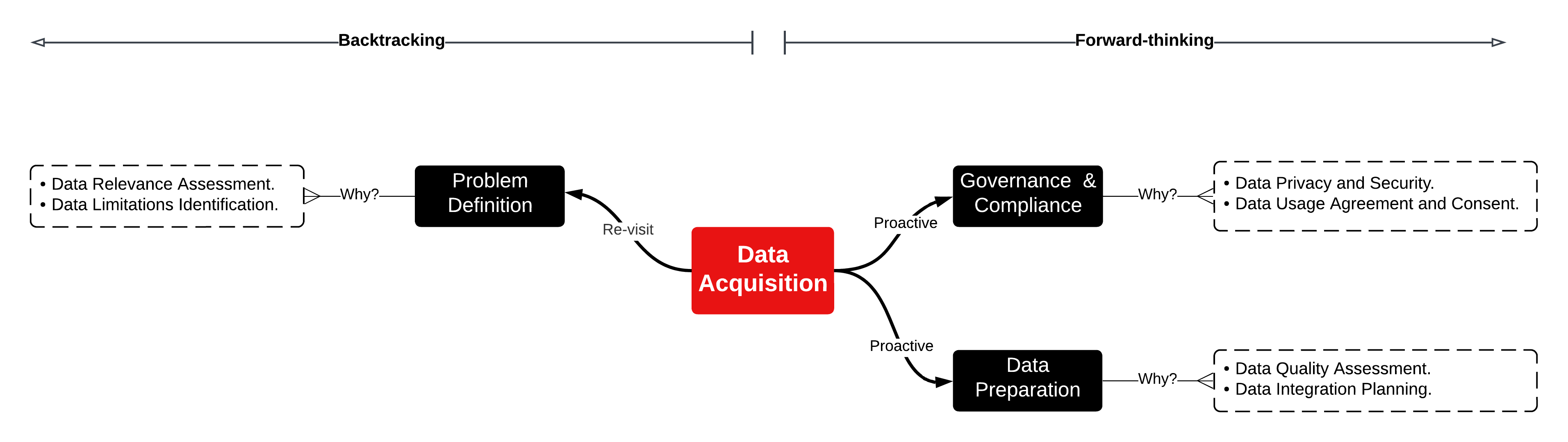}
		\caption{}\label{da2}
	\end{subfigure}
	\centering
	\caption{Main Focus: Data Acquisition  (a) The factual focus distribution when Data Acquisition phase is planned to be the main focus  (b) Data Acquisition phase inter-dependencies and main scenarios that trigger focus shift } \label{da}

\end{figure}

When focusing on data acquisition, it was observed that visits were common to three phases that are the problem definition, data preparation, and governance and compliance. These phases were essential for ensuring the quality and suitability of the acquired data, aligning it with the defined problem, and adhering to governance and compliance requirements as in Figure \ref{da}.

\subsubsection{Backtracking: Problem Definition}
Examples of scenarios:
\begin{enumerate}
	
	\item \textit{Data Relevance Assessment}: During the data acquisition process, you may come across new data sources or variables that raise questions about their relevance to the original problem definition. The need to reassess the relevance of the acquired data and its alignment with the problem definition arises. This may involve reevaluating the problem statement, objectives, and the data acquired to ensure they are aligned and address the intended problem accurately.
	
	\item  \textit{Data Limitations Identification}: While acquiring data, you may encounter limitations or constraints that affect the feasibility or scope of the problem definition. The need to identify and address these data limitations arises. For example, if the acquired data lacks certain variables or exhibits biases, it may impact the problem's feasibility or require adjustments in the problem formulation.

\end{enumerate}

\newpage
\subsubsection{Forward-thinking: Data Preparation} \label{pdata}
Examples of scenarios:

\begin{enumerate}
	
	\item \textit{Preliminary Data Quality Assessment}:  While acquiring data, it is essential to assess its quality to ensure it meets the desired standards for analysis and modeling. This may involve checking for missing values, outliers, or inconsistencies in the acquired data. Performing data quality assessment tasks, such as data cleaning or handling missing values, alongside data acquisition helps ensure that the acquired data is of high quality and ready for further analysis. \label{sc}
	
	\item  \textit{Data Integration Planning}: Data integration might have been touched to some extent during the problem definition phase, but during the data acquisition phase, it may become apparent that integrating additional data sources is necessary to enhance the overall dataset's completeness or predictive power. While primarily focused on acquiring data, it is important to simultaneously plan for the integration of these additional data sources. This includes identifying the integration process, mapping variables or features, and considering any data transformation or alignment required to integrate the datasets effectively, and most of that is under the data preparation phase scope.

\end{enumerate}

\subsubsection{Forward-thinking: Governance and Compliance}
Examples of scenarios:
\begin{enumerate}
	
	\item \textit{Data Privacy and Security Assessment}:  While acquiring data, it is crucial to assess the privacy and security aspects associated with the collected data. This includes evaluating whether the data acquisition process adheres to relevant data privacy regulations and policies. Conducting a privacy impact assessment and ensuring data security measures are in place should be done concurrently with data acquisition to safeguard sensitive information and maintain compliance with data protection regulations.
	
	\item  \textit{Data Usage Agreement and Consent}:  As part of governance and compliance, it is important to establish clear data usage agreements and obtain appropriate consent for the collected data. That is especially important when dealing with external stakeholders (e.g., external clients). While focusing on data acquisition, ensure that the necessary legal agreements, contracts, or consent forms are prepared and obtained from data providers or individuals whose data is being collected.

\end{enumerate}

\subsubsection {Recommendations for Optimizing the Data Acquisition Phase} 

To optimize the data acquisition phase, we recommend:

\begin{enumerate}
	\item Involving and collaborating with data experts, such as data engineers and infrastructure experts, early in the data acquisition process. Their expertise can help identify potential data challenges and assess data availability on data sources that align with your requirements. 
	
	\item Before acquiring data from external sources, perform due diligence on the data providers. Verify their credibility, data quality, and adherence to relevant regulations. Ensure that data providers have clear data usage policies and consent mechanisms in place.

	\item Considering to conduct a "pilot" data acquisition phase to gather a small sample of the data before committing to a full-scale acquisition. This allows you to assess the quality, relevance, and suitability of the acquired data early on.
	
	\item Maintain clear and comprehensive documentation of data sources, acquisition methods, and any limitations or caveats associated with the acquired data.
		
	\item Maintain detailed documentation of the data governance and compliance practices followed during the data acquisition process. This includes records of data provider due diligence, data usage agreements, anonymization techniques applied, and data privacy and security measures implemented.

\end{enumerate}

\newpage
\subsection{ Main Focus: Data Preparation}
\begin{figure}[htp]
	
	\begin{subfigure}{\textwidth}
		\centering
		\includegraphics[width=0.8\textwidth]{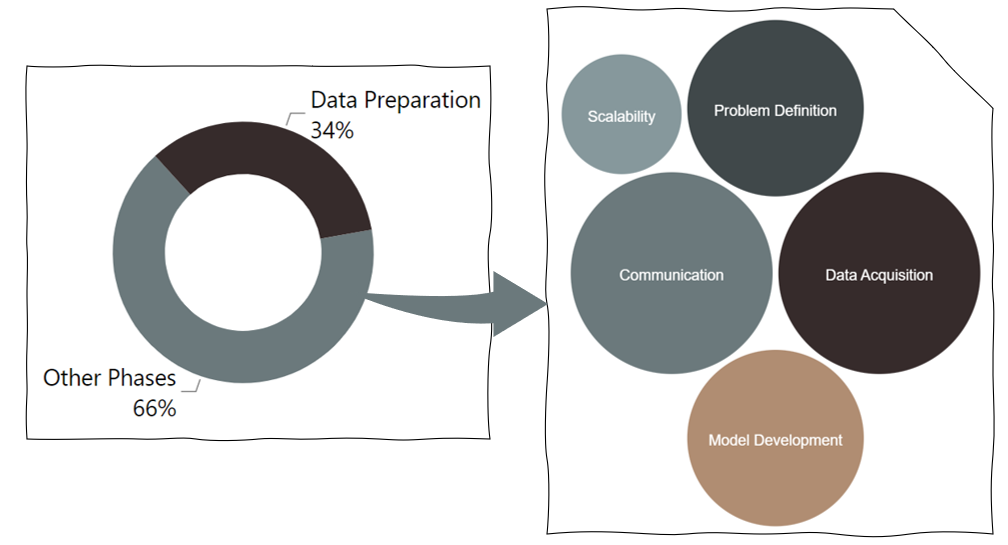}
		\caption{}\label{dp1}
	\end{subfigure}
	
	\bigskip
	
	\begin{subfigure}{\textwidth}
		\centering
		\includegraphics[width= \textwidth]{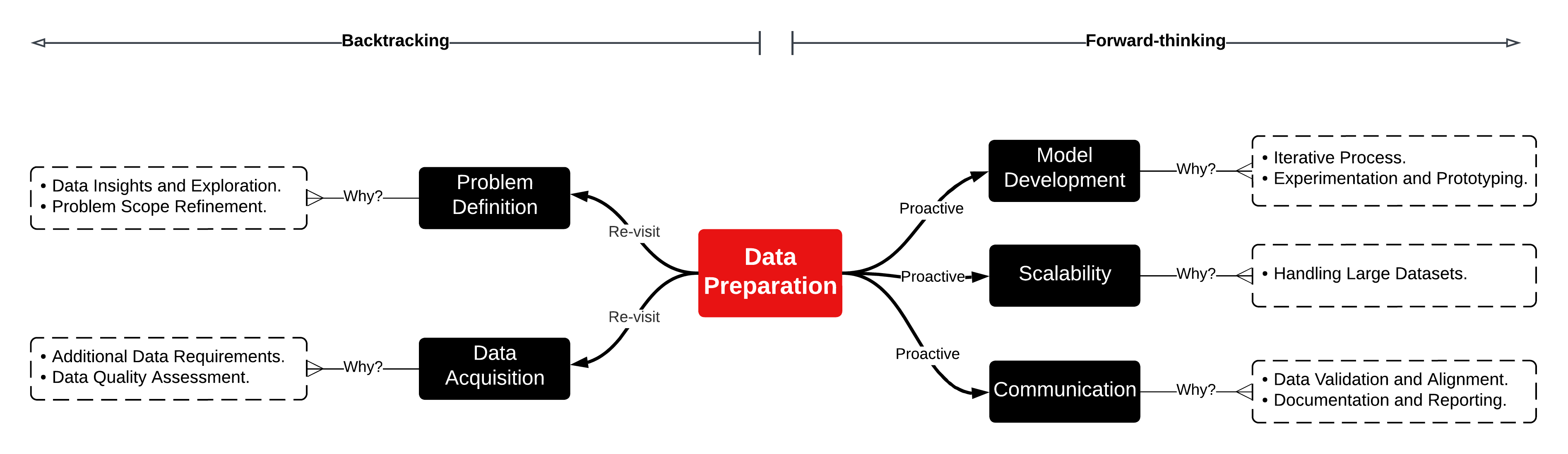}
		\caption{}\label{dp2}
	\end{subfigure}
	\centering
	\caption{Main Focus: Data Preparation  (a) The factual focus distribution when Data Preparation phase is planned to be the main focus  (b) Data Preparation phase inter-dependencies and main scenarios that trigger focus shift }\label{dp}

\end{figure}

When focusing on the data preparation phase, we found that it often involves visits to several other phases. These include back-tracking to the problem definition and data acquisition phases, as well as forward-thinking about the model  development and the scalability phases, and we need to carryout some tasks related to the communication phase as in Figure \ref{dp}. However, we observed less frequent visits to phases like model deployment, monitoring, and governance.

\subsubsection{Backtracking: Data Acquisition}

Examples of scenarios:
\begin{enumerate}
	
	\item \textit{Additional Data Requirements}: During the data preparation phase, you may identify the need for additional data or specific variables that were not initially acquired. This could involve identifying new data sources or exploring external data providers.
	
	\item  \textit{Data Quality Assessment}: This scenario pairs with  scenario \ref{sc} in section \ref{pdata} while focusing on the data acquisition phase, however that was a preliminary quality assessment. In the opposite direction, here, as part of the data preparation process, you may uncover data quality issues or limitations that cannot be adequately addressed within the current dataset. The need to reassess the data quality and explore alternative data sources arises, prompting a revisit to the data acquisition phase. This could involve seeking additional data to replace or supplement existing data with higher quality or more comprehensive sources.
	
\end{enumerate}

\subsubsection{Backtracking: Problem Definition}
Examples of scenarios:

\begin{enumerate} 
	\item  \textit{Problem Scope Refinement}: While preparing the data, you may discover limitations or complexities that affect the feasibility or scope of the problem definition. The need to reassess the problem statement arises to align it with the available data and its characteristics. This could involve adjusting the problem objectives or setting more realistic expectations based on the insights gained during the data preparation phase (e.g., focus on short-term customer churn instead of long-term churn).
	
	\item \textit{Data Insights and Exploration}: During the data preparation process, you may uncover valuable insights or patterns within the data that prompt a reassessment of the problem definition. The need to revisit the problem definition phase arises to incorporate these newfound insights into a more refined problem statement.

\end{enumerate}

\subsubsection{Forward-thinking: Model Development}
Examples of scenarios:

\begin{enumerate} 
	
	\item  \textit{Experimentation and Prototyping}: While the primary focus is on data preparation, there might be cases where you need to experiment with preliminary modeling approaches or develop prototypes. These tasks could involve training initial models on a subset of the prepared data to evaluate their performance or to analyze feature importance.

	\item \textit{Iterative Process}: Data preparation and model development are interconnected  and iterative. Challenges like skewness in features may require exploring alternative modeling techniques. Also, identifying domain-specific features during data preparation could lead to adjustments in the model development.

\end{enumerate}

\subsubsection{Forward-thinking: Scalability}
Data preparation phase itself does not have a direct dependency on the scalability phase. However, here is a scenario where there could be some interaction between the data preparation and scalability phases:

\begin{enumerate} 
	
	\item \textit{Handling Large Datasets}: In some cases, the data being prepared for modeling might be very large in size, requiring scalability considerations during the data preparation phase. If the dataset is too large to fit into memory or the processing time becomes prohibitively long, you may need to employ or think about scalable data processing techniques or distributed computing frameworks to handle the data effectively. You may discover at this early stage that you need to build cloud-based ML system to benefit from cloud scalability options.
	
\end{enumerate}

\subsubsection{Forward-thinking: Communication}
Examples of scenarios:

\begin{enumerate} 
	
	\item \textit{Data Validation and Alignment}: During the data preparation phase, effective communication with stakeholders and domain experts is important to validate and align the prepared data with the intended objectives of the ML project. By involving stakeholders in the data preparation process, you can ensure that the data is aligned with their expectations and that any specific considerations or constraints are taken into account. 
	
	\item  \textit{Documentation and Reporting}: Clear and concise documentation of the data preparation process, methodologies, and decisions made during this phase is essential. This documentation serves as a means of communication with other team members, including data scientists, engineers, or domain experts involved in subsequent phases. 
	
\end{enumerate}

\subsubsection {Recommendations for Optimizing the Data Preparation Phase} 

To optimize the data preparation phase, we recommend:

\begin{enumerate}
	
	\item Performing iterative data exploration and analysis during the data preparation phase to uncover valuable insights, patterns, and data quality issues. This helps refine the problem statement and align it with the available data.

	\item Engage stakeholders and domain experts throughout the data preparation phase to validate and align the prepared data with the intended objectives and requirements of the ML project. Their input helps ensure the data reflects their expectations and incorporates any specific considerations or constraints.
	
	\item Maintain clear and concise documentation of the data preparation process, methodologies, and decisions made. This documentation serves as a means of communication with team members involved in subsequent phases and facilitates seamless collaboration.

\end{enumerate}

\newpage
\subsection{ Main Focus: Model Development }

In this experiment, we found that, the model  development phase holds significant influence as it often drives decisions in future stages or introduces changes to previous stages see Figure \ref{mdev}. Here is how these inter-dependencies are:

\begin{figure}[htp]
	
	\begin{subfigure}{\textwidth}
		\centering
		\includegraphics[width=0.8\textwidth]{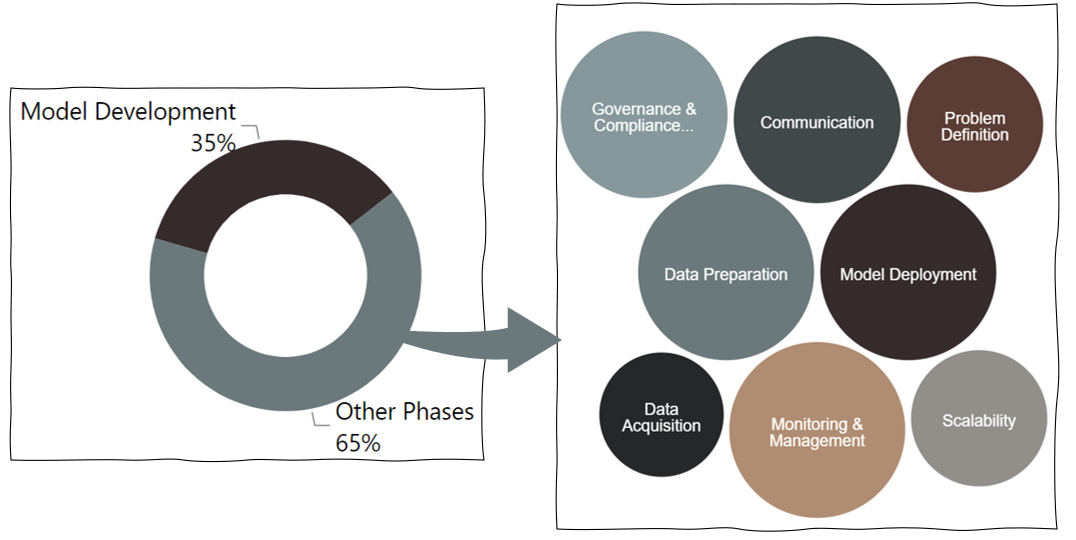}
		\caption{}\label{mdev1}
	\end{subfigure}
	
	\bigskip
	
	\begin{subfigure}{\textwidth}
		\centering
		\includegraphics[width= \textwidth]{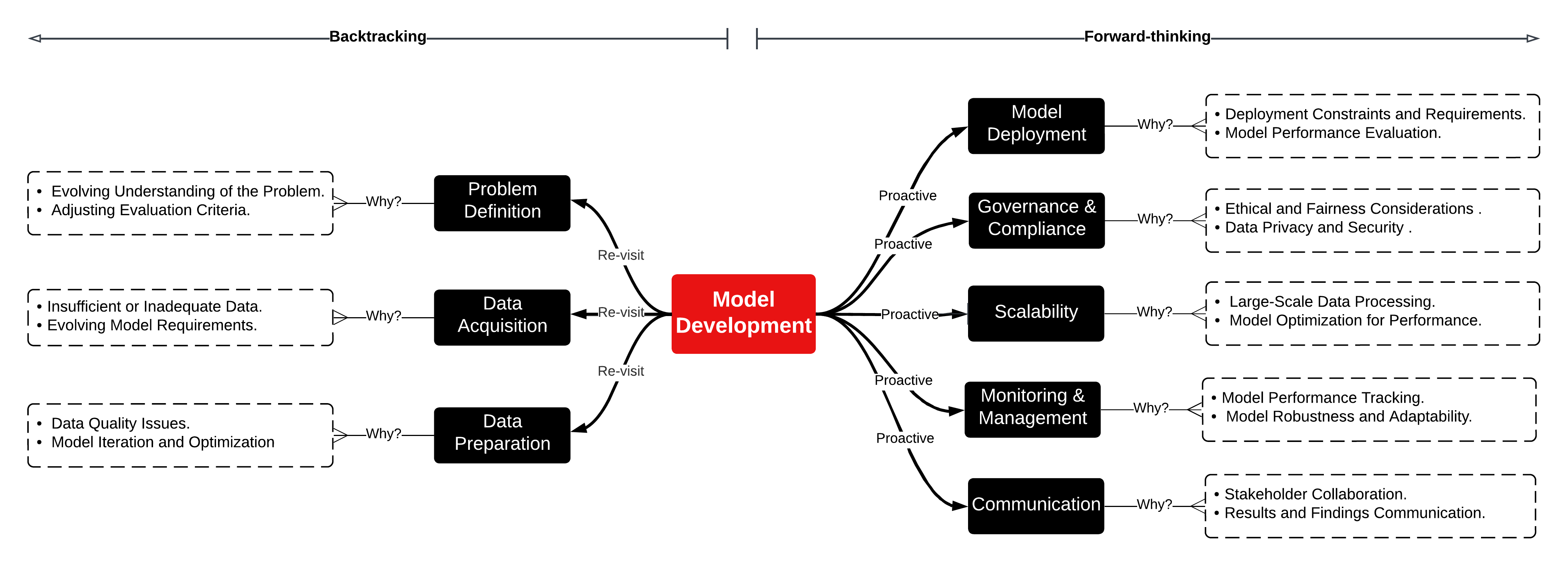}
		\caption{}\label{mdev2}
	\end{subfigure}
	\centering
	\caption{Main Focus:  Model Development   (a) The factual focus distribution when  Model Development  phase is planned to be the main focus  (b)  Model Development  phase inter-dependencies and main scenarios that trigger focus shift }\label{mdev} 
\end{figure}

\subsubsection{Backtracking: Problem Definition}
Examples of scenarios:

\begin{enumerate} 
	
	\item \textit{Evolving Understanding of the Problem}: As you delve deeper into the model  development phase, you may gain a better understanding of the problem and its nuances. This deeper understanding can lead to insights or challenges that require a re-evaluation of the initial problem definition. For example, you may discover additional factors that influence the problem, identify new subtasks or objectives, or uncover complexities that were not initially considered. 
	
	\item  \textit{Adjusting Performance Metrics and Evaluation Criteria}: During the model  development phase, you may find that the initially defined performance metrics or evaluation criteria are not capturing the most relevant aspects of the problem or are no longer suitable for the refined model objectives. 
	
\end{enumerate}

\subsubsection{Backtracking: Data Acquisition}
Examples of scenarios:
\begin{enumerate} 
	
	\item \textit{Insufficient or Inadequate Data}: During the model  development phase, you may realize that the data you initially acquired is insufficient or inadequate to train a model that meets your performance requirements. This could happen if the data lacks diversity, has imbalanced class distributions, or is missing important features. 
	
	\item  \textit{Evolving Model Requirements}: It is not unusual that as you progress in the model  development phase, you may  receive new insights from stakeholders that lead to changes in the model's requirements. These changes could involve the need for additional data attributes, specific data formats, or expanded data coverage.
	
\end{enumerate}

\subsubsection{Backtracking: Data Preparation}
Examples of scenarios:
\begin{enumerate} 
	
	\item \textit{Data Quality Issues}: During the model  development phase, you may discover data quality issues that were not adequately addressed during the initial data preparation phase. Addressing these data quality issues is crucial for model training, as poor-quality data can negatively impact the model's performance and generalization.
	
	\item  \textit{Model Iteration and Optimization}: Inevitably, model development involves an iterative process of training, evaluation, and refinement. For instance, in some cases, you may need to apply data augmentation techniques or an alternative word embedding  approach to improve your computer vision or NLP model performance. These types of adjustments are under the data preparation phase scope. 
\end{enumerate}

\subsubsection{Forward-thinking: Model Deployment}
While the main focus is on the model  development phase, there can be scenarios where it's beneficial to foresee the model deployment phase. Here are two scenarios that make model development phase dependent on the model deployment phase:

\begin{enumerate} 
	
	\item \textit{Deployment Constraints and Requirements}: Considering the deployment constraints and requirements during the model  development phase helps ensure that the developed model is compatible and can be smoothly integrated into the target deployment environment. By proactively thinking about the deployment phase, you can take into account factors such as hardware or software dependencies, latency or resource constraints, scalability requirements, or regulatory compliance. This enables you to design and develop the model with these considerations in mind, minimizing re-work later on.
	
	\item  \textit{Model Performance Evaluation}: Thinking about the model deployment phase during model  development allows you to consider how the model's performance will be monitored in the production environment. By anticipating the need for metrics, logging, or monitoring mechanisms during the development phase, you can  implement necessary hooks or instrumentation while designing and coding your model to facilitate effective monitoring and evaluation during deployment. 
	
\end{enumerate}

\subsubsection{Forward-thinking: Governance and Compliance}
Examples of scenarios:

\begin{enumerate} 
	
	\item  \textit{Data Privacy and Security}:  Thinking about data privacy and security considerations during the model  development phase helps protect sensitive information and comply with relevant regulations. Assessing data handling practices, access controls, and anonymization techniques can help safeguard user data and ensure compliance with data protection laws. 
		
	\item \textit{Ethical and Fairness Considerations}: Proactively considering ethical and fairness implications during the model  development phase ensures that the developed model aligns with regulatory requirements and ethical standards. By incorporating fairness metrics and assessing potential biases or discriminatory outcomes, you can identify and address any issues related to sensitive attributes, protected classes, or other fairness concerns.

\end{enumerate}

\subsubsection{Forward-thinking: Scalability}
Examples of scenarios:

\begin{enumerate} 
	
	\item  \textit{Model Optimization for Performance}: Thinking about scalability during the model  development phase helps optimize the model's performance when deployed in a scalable environment. By anticipating the need for efficient resource utilization and low-latency inference, you can design the model to optimize computational requirements and improve scalability. This might involve exploring techniques such as model pruning, quantization, or architecture optimization to reduce model complexity and enhance scalability. 
		
	\item \textit{Large-Scale Data Processing}: Considering scalability during the model  development phase becomes important when dealing with large-scale data processing. If you anticipate that the model will need to handle substantial volumes of data during deployment, it is crucial to design and develop the model with scalability in mind. This could involve using distributed computing frameworks, parallel processing techniques, or efficient data storage strategies.

\end{enumerate}

\subsubsection{Forward-thinking: Monitoring and Management}
Examples of scenarios:

\begin{enumerate} 
	
	\item \textit{Model Performance Tracking}: Considering monitoring and management during the model  development phase allows you to design the model with built-in mechanisms for tracking its performance. By incorporating appropriate logging and metrics during development, you enable effective monitoring of the model's behavior, predictions, and overall performance once it is deployed. 
	
	\item  \textit{Model Robustness and Adaptability}:  You might think beyond tracking and anticipate the need for implementing mechanisms for detecting data drift, data anomalies, or changes in performance over time. You better start this implementation or at least consider that while typing your code lines (e.g, add placeholders). This ensures a robust and adaptable model to changing conditions and it is under the scope of the monitoring and management phase. 
	
\end{enumerate}

\subsubsection{Forward-thinking: Communication}
Examples of scenarios:

\begin{enumerate} 
	
	\item \textit{Stakeholder Collaboration}: Effective collaboration with stakeholders is crucial during the model  development phase. Engaging in regular communication and feedback sessions with stakeholders allows you to gather insights, validate assumptions, and ensure that the developed model aligns with their expectations and requirements. You can address their concerns and clarify expectations.
	
	\item  \textit{Results and Findings Communication}:  You don't have to wait until the completion of the entire model  development phase and achieve an accurate model. It is beneficial to have intermediate checkpoints to share halfway results with stakeholders. These intermediate communication occasions help in setting benchmarks.
	
\end{enumerate}

\subsubsection {Recommendations for Optimizing the Model Development  Phase} 

To optimize the model  development phase, we recommend:

\begin{enumerate}
	\item Recognizing that model development is an iterative process. Plan for multiple iterations of training, evaluation, and refinement to gradually improve model performance. Implement mechanisms to capture insights and lessons learned from each iteration to inform subsequent improvements.
	
	\item  As you delve deeper into the model  development phase, gain a better understanding of the problem and its nuances. Revisit the problem definition phase to refine and update the problem statement, objectives, and evaluation criteria based on this evolving understanding.
	
	\item Identifying and address data limitations, such as insufficient or inadequate data, during the model  development phase. Revisit the data acquisition phase to acquire additional relevant data or modify the data collection strategy to meet the evolving model requirements.
	
	\item Proactively considering deployment constraints, requirements, and scalability during the model  development phase. Think about compatibility with the deployment environment, resource constraints, and scalability needs to design and develop the model with these considerations in mind.
	
	\item Considering ethical and fairness implications during the model  development phase. Incorporate fairness metrics, assess potential biases, and ensure compliance with regulatory requirements to develop models that align with governance and fairness guidelines.
	
	\item Thinking about monitoring and management considerations during model  development. Incorporate mechanisms for tracking model performance, detecting  drifts, and addressing data anomalies to facilitate effective monitoring and model management during deployment.

\end{enumerate}

\newpage
\subsection{ Main Focus: Model Deployment }

\begin{figure}[htp]
	
	\begin{subfigure}{\textwidth}
		\centering
		\includegraphics[width=0.8\textwidth]{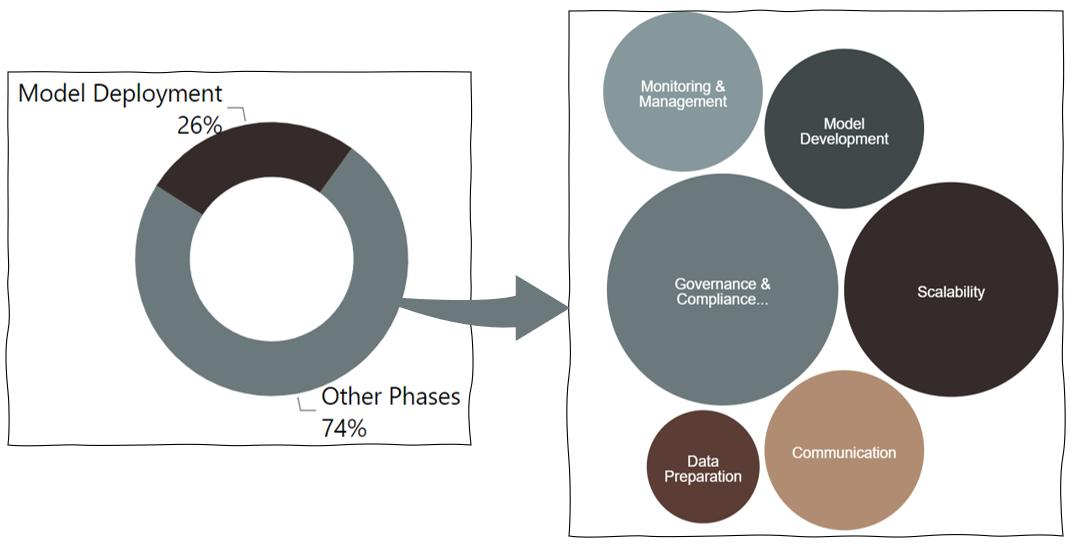}
		\caption{}\label{mdep1}
	\end{subfigure}
	
	\bigskip
	
	\begin{subfigure}{\textwidth}
		\centering
		\includegraphics[width= \textwidth]{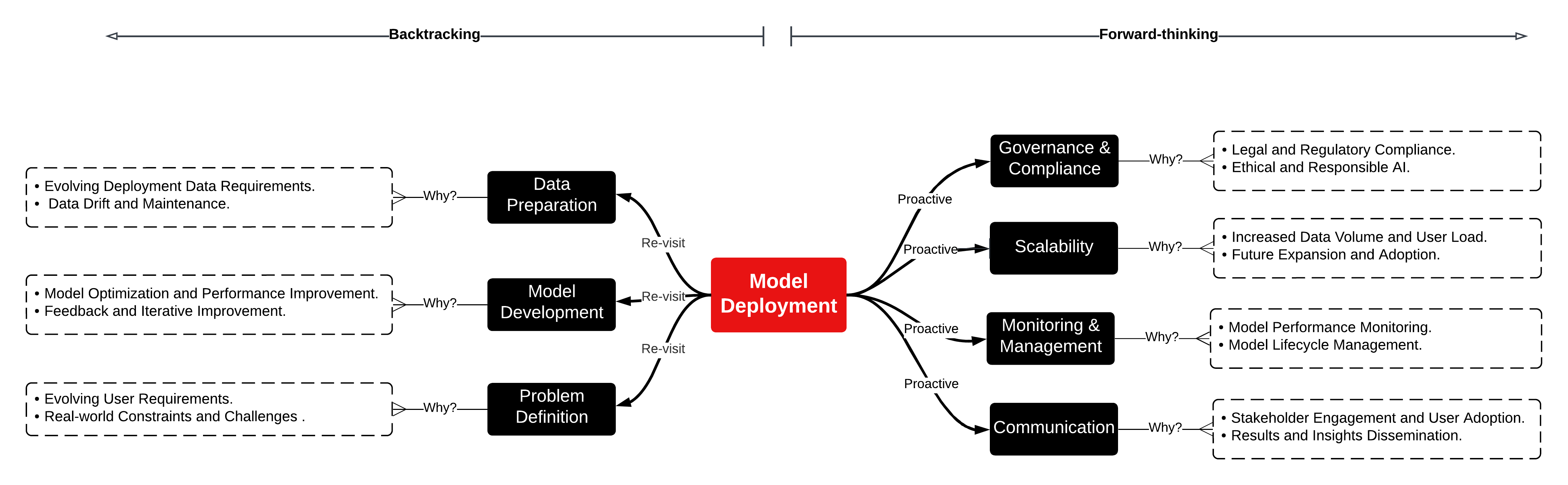}
		\caption{}\label{mdep2}
	\end{subfigure}
	\centering
	\caption{Main Focus:  Model Deployment   (a) The factual focus distribution when  Model Deployment  phase is planned to be the main focus  (b)  Model Deployment  phase inter-dependencies and main scenarios that trigger focus shift } \label{mdep}
\end{figure}

Results emphasize the close relationships and dependencies between the model deployment phase and various other phases, including governance and compliance, scalability, Monitoring, and model  development, as in Figure \ref{mdep}.

\subsubsection{Backtracking: Problem Definition}
Examples of scenarios:

\begin{enumerate} 

	\item  \textit{Real-world Constraints and Integration Challenges}:  During model deployment, unexpected real-world constraints and integration challenges may arise, such as technical limitations, resource constraints, security considerations, or operational costs. For instance, in a real-time object detection model for autonomous vehicles, a problem definition change may involve adjusting the statement to process lower-resolution data from a single sensor. 
	
\end{enumerate}

\subsubsection{Backtracking: Data Preparation}
Examples of scenarios:

\begin{enumerate} 
	
	\item \textit{Evolving Deployment Data Requirements}: As you move into the model deployment phase, you may realize that the data requirements for deployment differ from those initially considered during the data preparation phase. Deployment environments might have specific data format or integration requirements that were not accounted for previously. Revisiting the data preparation phase allows you to re-evaluate and adjust the prepared data to meet the specific requirements of the deployment environment.  
	
	\item  \textit{Data Drift and Maintenance}:  Model development is a time-consuming process, and it is common practice to save a training dataset during this phase. However, this dataset may become stale by the time we deploy the model. If the model development process takes an extended period and the data drifts, deploying the model could reveal the need to review and update the data preparation pipelines.
	
\end{enumerate}

\subsubsection{Backtracking: Model Development}
Examples of scenarios:

\begin{enumerate}

	\item  \textit{Feedback and Iterative Improvement}: During the model deployment phase, you may receive feedback from users, stakeholders, or monitoring systems that highlight areas where the deployed model can be further improved. This feedback might uncover use cases or scenarios that were not fully anticipated during the initial model development. Revisiting the model  development phase enables you to incorporate this feedback into the development process, iteratively improving the model. 
	
\end{enumerate}

\subsubsection{Forward-thinking: Monitoring and Management }
Examples of scenarios:

\begin{enumerate} 
	
	\item \textit{Model Performance Monitoring}: Thinking about monitoring during the model deployment phase is practical consideration allows you to design and implement mechanisms for tracking and evaluating the model's performance in the production environment. You can plan for the collection and analysis of relevant performance metrics, real-time inference logs, and error tracking. 
	
	\item  \textit{Model Maintenance and Lifecycle Management}: Thinking about management during the model deployment phase enables you to proactively plan for ongoing model maintenance and lifecycle management. This includes tasks such as version control, model updates, retraining schedules, and documentation. Proactively thinking about management helps ensure the long-term viability of the deployed model.
	
\end{enumerate}

\subsubsection{Forward-thinking: Scalability }
Examples of scenarios:

\begin{enumerate} 
	
	\item \textit{Increased Data Volume and User Load}: As you progress in the model deployment phase, the volume of data and user load may increase beyond the initial expectations. This growth can pose challenges in terms of processing capacity, response time, and resource utilization. By proactively considering scalability, you can design the deployment infrastructure to handle the anticipated increase in data volume and user load. This may involve leveraging scalable computing resources, implementing load balancing mechanisms, or using distributed processing frameworks. 
	
	\item  \textit{Future Expansion and Adoption}: If there is potential for the model to be utilized by a larger user base, integrated into additional systems, or deployed across multiple locations, scalability considerations become essential. By anticipating future needs, you can design the deployment architecture and infrastructure to be easily scalable and adaptable. 
	
\end{enumerate}

\subsubsection{Forward-thinking: Governance and Compliance }

Examples of scenarios:
\begin{enumerate} 
	
	\item \textit{Legal and Regulatory Compliance}: Considering governance and compliance during the model deployment phase helps ensure that the deployed model adheres to relevant legal and regulatory requirements. Different jurisdictions may have specific regulations regarding data privacy and security.

\end{enumerate}

\subsubsection{Forward-thinking: Communication }

Examples of scenarios:
\begin{enumerate} 
	
	\item \textit{Stakeholder Engagement and User Adoption}: Communication during the model deployment phase allows you to plan for effective stakeholder engagement and user adoption of the deployed model. You can develop strategies for conveying the benefits, limitations, and appropriate use cases of the model to stakeholders, users, and other relevant parties.

\end{enumerate}

\subsubsection {Recommendations for Optimizing the Model Deployment Phase} 

To optimize the model deployment phase, we recommend

\begin{enumerate}
	\item Ensuring that the problem definition is well-adapted to real-world constraints, and ensure that the deployed model addresses the most current needs.
	
	\item To consider monitoring and management requirements during model deployment. Design mechanisms for tracking model performance, detecting data drift, and managing the model's lifecycle. 
	
	\item To think ahead about scalability considerations to accommodate increased data volume, user load, and future expansion. Design the deployment infrastructure to handle growth without compromising performance, leveraging scalable computing resources and distributed processing frameworks.
	
	\item Ensure compliance with legal, regulatory, and ethical requirements during model deployment. Proactively design the deployment process and model architecture to meet privacy, security, fairness, and transparency guidelines.

\end{enumerate}

\newpage
\subsection{ Main Focus: Monitoring and Management}

The monitoring and management phase is closely interconnected and often overlaps with model  development, model deployment, and governance and compliance. Additionally, effective communication plays an important role in this phase.

\begin{figure}[htp]
	
	\begin{subfigure}{\textwidth}
		\centering
		\includegraphics[width=0.8\textwidth]{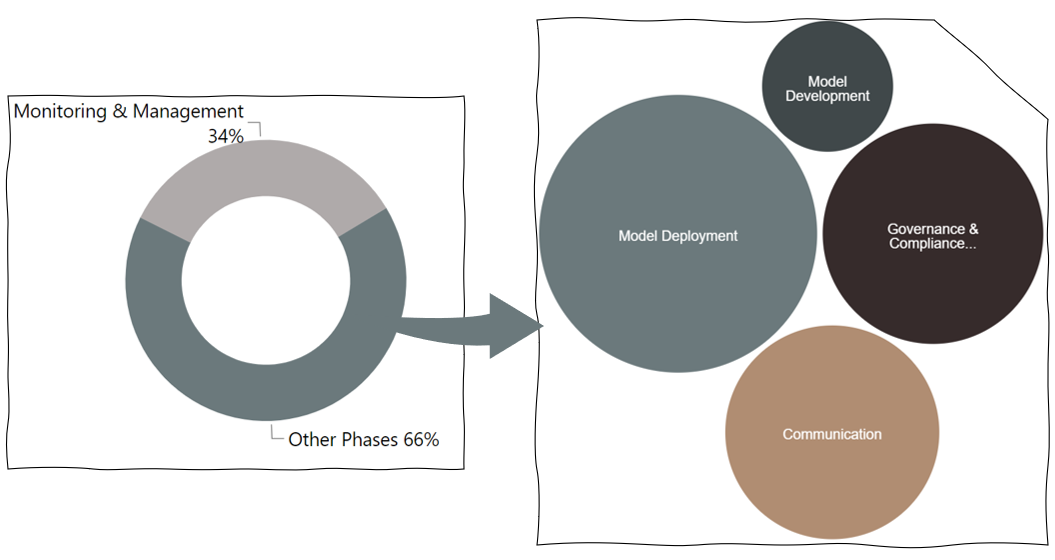}
		\caption{}\label{mon_}
	\end{subfigure}
	
	\bigskip
	
	\begin{subfigure}{\textwidth}
		\centering
		\includegraphics[width= \textwidth]{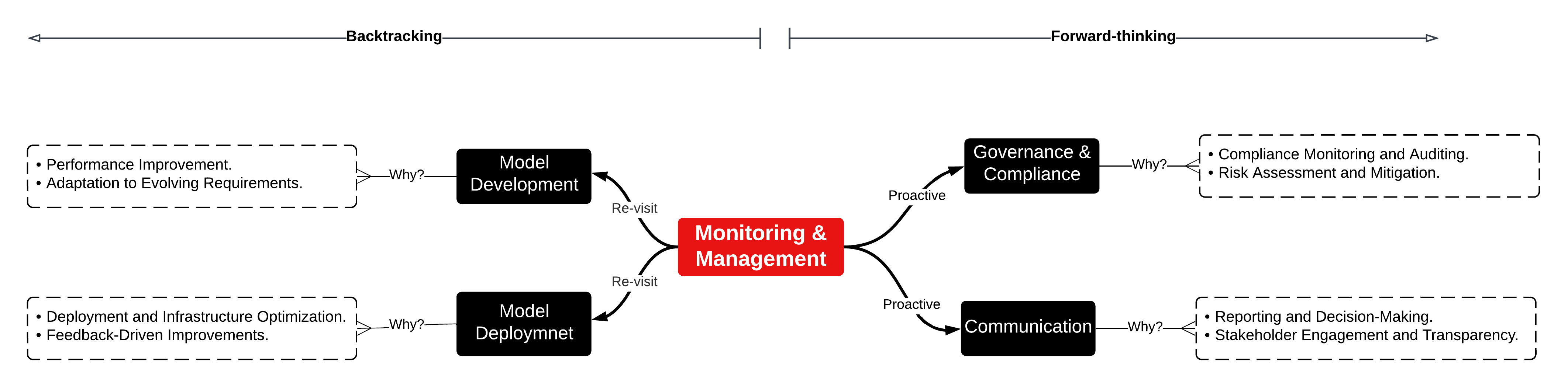}
		\caption{}\label{mon}
	\end{subfigure}
	\centering
	\caption{Main Focus:  Monitoring and Management   (a) The factual focus distribution when  Monitoring and Management  phase is planned to be the main focus  (b)  Monitoring and Management phase inter-dependencies and main scenarios that trigger focus shift }
\end{figure}

\subsubsection{Backtracking: Model Development }

Examples of scenarios:

\begin{enumerate} 
	
	\item \textit{Performance Improvement and Iterative Development}: During the monitoring and management phase, insights gained from monitoring the deployed model may reveal opportunities for performance improvement or the need for model updates. You may need to revisit the model  development phase to re-evaluate the model architecture, algorithms, or training data to address any identified issues or areas for enhancement. 
	
	\item  \textit{Adaptation to Evolving Requirements}:  The monitoring and management phase may highlight changing user needs, emerging use cases, or shifts in the business environment. Revisiting the model  development phase allows you to adapt the deployed model to these evolving requirements.

\end{enumerate}

\subsubsection{Backtracking: Model Deployment}

Examples of scenarios:

\begin{enumerate} 
	
	\item \textit{Deployment Optimization and Infrastructure Changes}: During the monitoring and management phase, you may identify opportunities to optimize the deployment process or make infrastructure changes to better support monitoring and management activities. This could include modifying the deployment architecture, updating deployment pipelines, or integrating additional monitoring tools or systems.

\end{enumerate}

\subsubsection{Forward-thinking: Governance and Compliance }

Examples of scenarios:

\begin{enumerate} 
	
	\item \textit{Compliance Monitoring and Auditing}: To allow you to establish mechanisms for monitoring and auditing the deployed model's compliance with relevant regulations, policies, and ethical guidelines. By thinking ahead about governance and compliance, you can design monitoring processes, data logging, and tracking mechanisms to ensure ongoing compliance. 
	
	\item  \textit{Risk Assessment and Mitigation}:  Thinking about governance and compliance during the monitoring and management phase enables you to proactively assess and mitigate risks associated with the deployed model. This includes risks related to data privacy, security, fairness, transparency, or unintended consequences.

\end{enumerate}

\subsubsection{Forward-thinking: Communication }

Examples of scenarios:

\begin{enumerate} 

	\item  \textit{Stakeholder Engagement and Transparency}:  Communication during the monitoring and management phase enables you to proactively plan for stakeholder engagement and ensure transparency regarding the monitoring process and outcomes.

\end{enumerate}

\subsubsection {Recommendations for Optimizing the Monitoring and Management Phase} 

To optimize the monitoring and management phase, we recommend:

\begin{enumerate}
	\item Ensuring thorough planning for the monitoring and management phase at the beginning of the project. Clearly define the key performance indicators (KPIs) and monitoring metrics that align with the project's objectives and stakeholder requirements.
	
	\item Adopting a culture of continuous integration and collaboration between the model  development and Monitoring phases. Ensure that the monitoring systems are seamlessly integrated into the deployment pipeline.
	
\end{enumerate}

\newpage
\subsection{ Main Focus: Scalability }

\begin{figure}[htp]
	
	\begin{subfigure}{\textwidth}
		\centering
		\includegraphics[width=0.8\textwidth]{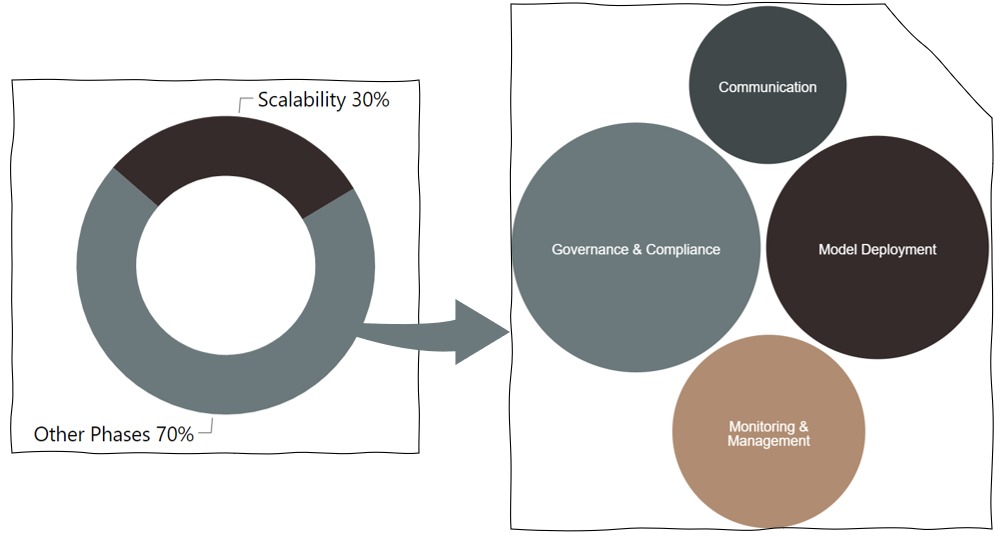}
		\caption{}\label{sca1}
	\end{subfigure}
	
	\bigskip
	
	\begin{subfigure}{\textwidth}
		\centering
		\includegraphics[width= \textwidth]{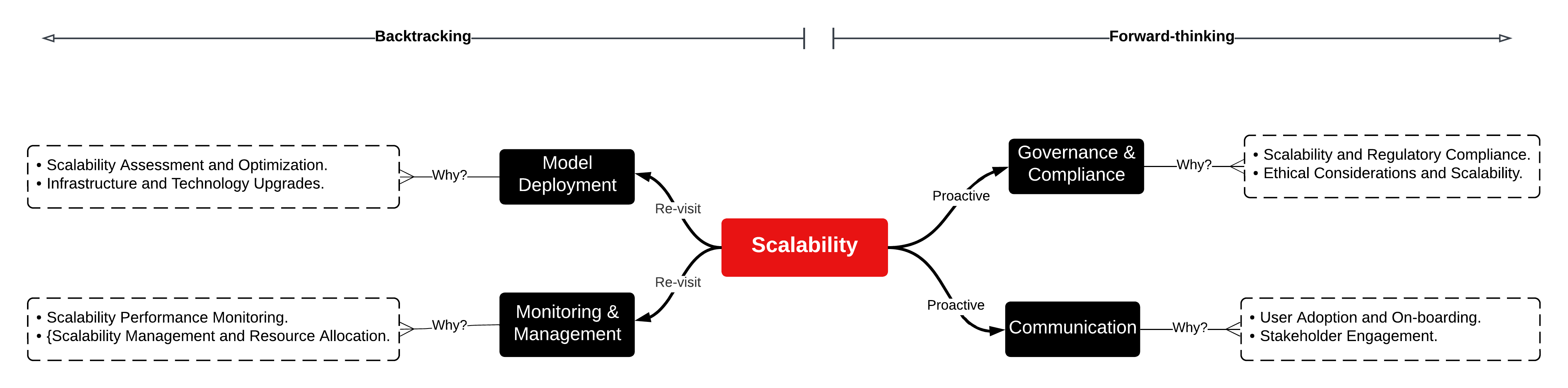}
		\caption{}\label{sca2}
	\end{subfigure}
	\centering
	\caption{Main Focus:  Scalability   (a) The factual focus distribution when  Scalability  phase is planned to be the main focus  (b)  Scalability phase inter-dependencies and main scenarios that trigger focus shift } \label{sca}
\end{figure}

During the scalability phase, four other phases are considered with significant overlap: governance and compliance, model deployment, monitoring and management, and communication as in Figure \ref{sca}.

\subsubsection{Backtracking: Model Deployment}

Examples of scenarios:

\begin{enumerate} 
	
	\item \textit{Scalability Assessment and Optimization}: When systematically addressing scalability phase you will evaluate the model's performance under increasing workloads, measuring resource utilization, or identifying potential bottlenecks. This may make you revisiting the model deployment phase, so you can make changes or optimizations to enhance the model's scalability. This may include modifying the deployment architecture, optimizing resource allocation, or implementing distributed computing techniques. 
	
	\item  \textit{Infrastructure and Technology Upgrades: } As you scale the deployment, you may encounter the need for infrastructure upgrades or technology changes to support the increased demand. This could involve migrating to more robust hardware or cloud-based infrastructure, adopting containerization technologies, or implementing scalable data storage solutions. 
	
\end{enumerate}

\subsubsection{Backtracking: Monitoring and Management}

Examples of scenarios:

\begin{enumerate} 
	
	\item \textit{Scalability Performance Monitoring}: During the scalability phase, as you scale the deployment to handle larger workloads or accommodate increased data volume, it becomes crucial to monitor the performance of the deployed model under these new conditions. This involves assessing how the model performs in terms of response times, resource utilization, and overall system performance as the workload increases. You may need to adjust the monitoring processes to effectively capture and analyze the scalability performance of the model. 
	
	\item  \textit{Resource Allocation and Management: } As the deployment scales, it may require efficient resource allocation and management to ensure optimal performance and cost-effectiveness. This includes dynamically allocating resources, such as computational power, memory, or storage, based on workload demands and scaling requirements.

\end{enumerate}

\subsubsection{Forward-thinking: Communication }

Examples of scenarios:

\begin{enumerate} 
	
	\item \textit{Stakeholder Engagement}: Expanding the user base or accommodating a larger audience requires effective communication with stakeholders. The goal of this communication is to inform them about scalability improvements, updates, and any changes that may affect their experience.
	
	\item  \textit{User Adoption and On-boarding: } Scalability means a broader user base. You need to ensure smooth user adoption and onboarding processes. You should effectively communicate the benefits of the scalable deployment and address user concerns.

\end{enumerate}

\subsubsection{Forward-thinking: Governance and Compliance  }

Examples of scenarios:

\begin{enumerate} 
	
	\item \textit{Scalability and Regulatory Compliance}: As you scale the deployment, it is crucial to ensure that the scalability efforts align with regulatory requirements and adhere to relevant governance policies. This includes considerations such as data privacy and security. This might involve reviewing and updating policies, conducting risk assessments, or implementing controls to ensure that the scalable deployment remains compliant with regulatory obligations.
	
	\item  \textit{Ethical Considerations and Scalability: } Scalability can introduce new ethical challenges that need to be addressed. As the user base expands and the deployment handles larger volumes of data, it becomes important to revisit ethical considerations to ensure fair and unbiased treatment, transparency, and accountability. 
	
\end{enumerate}

\subsubsection {Recommendations for Optimizing the Scalability Phase} 

To optimize the scalability phase, we recommend:

\begin{enumerate}
	\item Conducting a comprehensive scalability assessment during the early phases of development to Identify potential scalability challenges and plan for future growth and increased demand. This proactive approach helps you design a scalable architecture from the outset.
	
	\item Designing the deployment architecture to be modular and distributed, enabling seamless scaling by adding or removing resources as needed. Leverage technologies such as containerization and orchestration platform to enhance scalability.
	
	\item Conducing thorough load testing to simulate various usage scenarios and identifying performance bottlenecks. Use the insights gained to optimize the model's performance and resource utilization.
	
	\item Utilizing cloud-based infrastructure and services that offer elasticity and scalability. Cloud providers can automatically scale resources based on demand, reducing the need for manual adjustments.

\end{enumerate}

\newpage
\subsection{ Main Focus: Governance and Compliance}

During the governance and compliance phase, several other phases are considered with significant overlap. These include the problem definition phase, data acquisition phase, data preparation phase, model  development phase, model deployment as well as the importance of communication.

\begin{figure}[htp]
	
	\begin{subfigure}{\textwidth}
		\centering
		\includegraphics[width=0.8\textwidth]{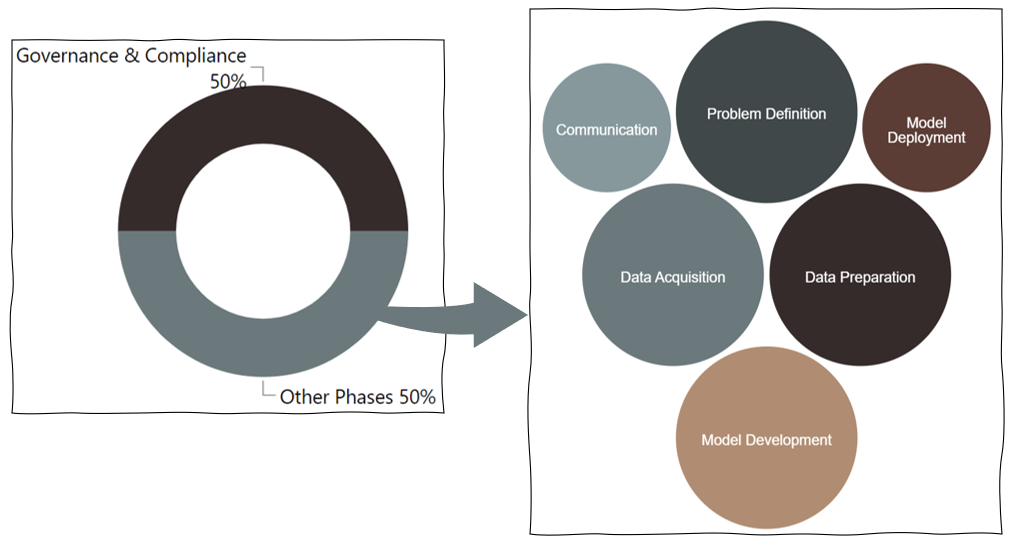}
		\caption{}\label{gov_}
	\end{subfigure}
	
	\bigskip
	
	\begin{subfigure}{\textwidth}
		\centering
		\includegraphics[width= \textwidth]{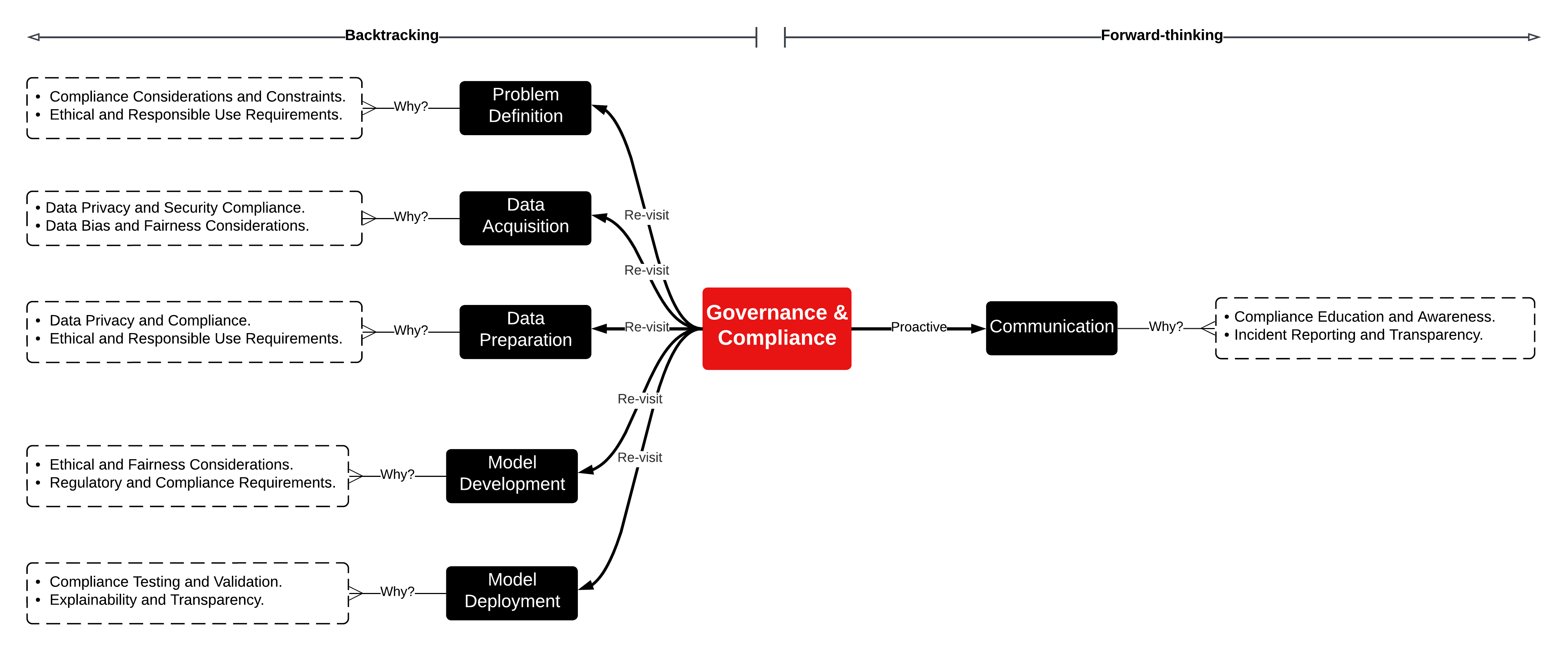}
		\caption{}\label{gov}
	\end{subfigure}
	\centering
	\caption{Main Focus:  Governance and Compliance   (a) The factual focus distribution when  Governance and Compliance  phase is planned to be the main focus  (b)  Governance and Compliance phase inter-dependencies and main scenarios that trigger focus shift }
\end{figure}

\subsubsection{Backtracking: Problem Definition }

Examples of scenarios:

\begin{enumerate} 
	
	\item \textit{Compliance Considerations and Constraints}: Covernance and compliance efforts require aligning the model development and deployment process with relevant regulations, policies, and ethical guidelines. During the governance and compliance phase, it may become apparent that certain compliance considerations or constraints were not adequately addressed in the initial problem definition phase. In such cases, revisiting the problem definition phase allows for a review of the problem statement, objectives, or requirements to ensure they align with the desired governance and compliance frameworks. 
	
	\item  \textit{Ethical and Responsible Use Requirements}: Governance and Compliance efforts emphasize the importance of ethical and responsible use of AI models. During the governance and compliance phase, it may be identified that the initial problem definition did not sufficiently consider ethical or responsible use requirements. You may need to reassess the problem formulation, biases, fairness considerations, or risks associated with the intended use of the model. 
	
\end{enumerate}

\subsubsection{Backtracking: Data Acquisition  }

Examples of scenarios:

\begin{enumerate} 
	
	\item \textit{Data Privacy and Security Compliance}: During the governance and compliance phase, it may become evident that the initial data acquisition process did not adequately address certain privacy or security requirements. You may need to review of the data collection practices, consent mechanisms or data storage protocols to ensure compliance.
	
	\item  \textit{Data Bias and Fairness Considerations: } It may be identified that the data used for model development introduces bias or lacks diversity, potentially leading to unfair treatment or biased outcomes. 
	
\end{enumerate}

\subsubsection{Backtracking: Data Preparation }

Examples of scenarios:

\begin{enumerate} 
	
	\item \textit{Data Privacy and Compliance}: Although privacy and security considerations may have been taken into account during data preparation, the governance phase focuses on a systematic assessment. In some cases, it becomes apparent that the initial data preparation processes did not fully address specific privacy or compliance requirements. As a result, a review of the data preparation may be necessary during this phase. 
	
	\item  \textit{Ethical and Responsible Use Requirements}: Similar to the privacy compliance, while you may have taken into account ethical and responsible use requirements during data preparation, the governance phase involves a systematic assessment. In certain instances, it becomes evident that the initial data preparation processes did not sufficiently address specific ethical and responsible use considerations. As a result, a review and potential adjustments during the governance phase are necessary to ensure full compliance with these requirements.

\end{enumerate}

\subsubsection{Backtracking: Model Development  }

Examples of scenarios:

\begin{enumerate} 
	
	\item \textit{Ethical and Fairness Considerations}: The governance phase involves a systematic assessment to ensure compliance with ethical and fairness considerations. During this phase, a review and potential adjustments are necessary to address any instances where the initial model development processes did not adequately account for specific ethical and fairness considerations, even though they may have been considered during the model development.
	
	\item  \textit{Regulatory and Compliance Requirements } During the governance phase, we carefully assess to ensure we meet all regulatory and compliance requirements. This means reviewing our model development processes and making any necessary adjustments. Even though we may have considered these requirements during the model development, the governance phase helps us systematically assess and address any gaps in compliance, making sure we meet all necessary regulations.
	
\end{enumerate}

\subsubsection{Backtracking: Model Deployment  }

Examples of scenarios:

\begin{enumerate} 
	
	\item \textit{Compliance Testing and Validation}: Governance and Compliance efforts often involve testing and validating the deployed model to ensure its adherence to relevant regulations, policies, and ethical guidelines. Similarly, it may be necessary to review and reassess the model deployment process to ensure compliance with these requirements.

\end{enumerate}

\subsubsection{Forward-thinking: Communication }

Examples of scenarios:

\begin{enumerate} 
	
	\item \textit{Compliance Education and Awareness}: This phase ideally involves educating stakeholders, users, and decision-makers about the regulatory requirements and ethical guidelines surrounding the deployed model. It may be necessary to  enhance the communication strategies to effectively convey compliance-related information. This could include providing clear guidelines, policies, or training materials that inform stakeholders about their responsibilities and obligations.
	
	\item  \textit{Incident Reporting and Transparency}: This phase emphasizes the importance of incident reporting, transparency, and accountability in the model deployment process.  You need to communicate  address incident reporting procedures and ensure transparency in communicating any breaches or non-compliance issues. 
	
\end{enumerate}

\subsubsection {Recommendations for Optimizing the Governance and Compliance Phase} 

To optimize the governance and compliance phase, we recommend:

\begin{enumerate}
	\item  Integrating governance and compliance considerations into the early phases of the project, including problem definition, data acquisition, and model development. This ensures that governance requirements are addressed from the outset and reduces the need for backtracking to previous phases.
	
	\item Conducting thorough research on relevant regulations, policies, and ethical guidelines that apply to the project. Stay up-to-date with any changes or updates to ensure ongoing compliance.
	
	\item Establishing an ethics review board or committee-like to provide ongoing oversight and guidance on ethical considerations, including fairness, bias, and transparency.

\end{enumerate}

\newpage
\section{Conclusion}

In the context of MLOps, we have emphasized the distinction between iterativeness and repetitiveness, advocating for an iterative approach where each repetition in the process aims for improvement and refinement.

To optimize the entire workflow and enhance efficiency while reducing frustrations, a collective and thoughtful approach to the stages is vital. Understanding the interdependencies and iterative characteristics of the MLOps process allows us to strategically address overlaps and streamline the workflow, resulting in improved outcomes for machine learning projects.

Throughout this article, practical recommendations have been provided for each MLOps phase, from problem definition to governance and compliance, underscoring the importance of continuous engagement, collaboration, and proactive planning.

The optimization of the MLOps workflow necessitates a proactive mindset, a commitment to continuous improvement, and an understanding of evolving problems. By aligning efforts with an iterative approach and avoiding repetitiveness, we can confidently navigate the dynamic landscape of MLOps, ensuring efficient deployment of AI solutions and maximizing the value of machine learning in real-world applications.

\section*{References}

{
\small

[1] Huyen, C., 2022. Designing machine learning systems. " O'Reilly Media, Inc.".

%%%%%%%%%%%%%%%%%%%%%%%%%%%%%%%%%%%%%%%%%%%%%%%%%%%%%%%%%%%%

\end{document}